\documentclass[
reprint,
nofootinbib,
 amsmath,amssymb,
 aps,
pra,
superscriptaddress,
]{revtex4-2}

\usepackage{graphicx}
\usepackage{dcolumn}
\usepackage{bm}

\newcommand{\xe}{\tilde{X}(300)}
\newcommand{\xg}{\tilde{X}(000)}
\newcommand{\Ae}{\tilde{A}(100)}
\newcommand{\ag}{\tilde{A}(000)}

\newcommand{\CaltechAPh}{Division of Engineering and Applied Science, California Institute of Technology, Pasadena, CA 91125, USA}

\newcommand{\CaltechPh}{Division of Physics, Mathematics, and Astronomy, California Institute of Technology, Pasadena, CA 91125, USA}

\begin{document}

\title{Direct measurement of high-lying vibrational repumping transitions for molecular laser cooling}

 \author{Nickolas H. Pilgram}
 \altaffiliation[Current affiliation: ]{Sensor Sciences Division, National Institute of Standards and Technology, Gaithersburg, MD 20899, USA}
\affiliation{\CaltechAPh}

 \author{Arian Jadbabaie}
\affiliation{\CaltechPh}

\author{Chandler J. Conn}
\affiliation{\CaltechPh}

\author{Nicholas R. Hutzler}
\affiliation{\CaltechPh}

\date{\today}

\begin{abstract}
Molecular laser cooling and trapping requires addressing all spontaneous decays to excited vibrational states that occur at the $\gtrsim 10^{-4} - 10^{-5}$ level, which is accomplished by driving repumping transitions out of these states. However, the transitions must first be identified spectroscopically at high-resolution.  A typical approach is to prepare molecules in excited vibrational states via optical cycling and pumping, which requires multiple high-power lasers. Here, we demonstrate a general method to perform this spectroscopy without the need for optical cycling. We produce molecules in excited vibrational states by using optically-driven chemical reactions in a cryogenic buffer gas cell, and implement frequency-modulated absorption to perform direct, sensitive, high-resolution spectroscopy. We demonstrate this technique by measuring the spectrum of the $\tilde{A}^2\Pi_{1/2}(1,0,0)-\tilde{X}^2\Sigma^+(3,0,0)$ band in $^{174}$YbOH. We identify the specific vibrational repump transitions needed for photon cycling, and combine our data with previous measurements of the $\tilde{A}^2\Pi_{1/2}(1,0,0)-\tilde{X}^2\Sigma^+(0,0,0)$ band to determine all of the relevant spectral constants of the $\tilde{X}^2\Sigma^+(3,0,0)$ state.  This technique achieves high signal-to-noise, can be further improved to measure increasingly high-lying vibrational states, and is applicable to other molecular species favorable for laser cooling.

\end{abstract}

\maketitle

\section{\label{sec:Intro}Introduction}

Molecules have unique applications in precision measurements of fundamental physics~\cite{safronovaSearchNewPhysics2018,hutzlerPolyatomicMoleculesQuantum2020}, quantum information processing \cite{demilleQuantumComputationTrapped2002,niDipolarExchangeQuantum2018,yuScalableQuantumComputing2019,albertRobustEncodingQubit2020}, quantum simulation \cite{bohnColdMoleculesProgress2017,mosesNewFrontiersQuantum2017,blackmoreUltracoldMoleculesQuantum2018,wallRealizingUnconventionalQuantum2015}, and controlled quantum chemistry \cite{balakrishnanPerspectiveUltracoldMolecules2016,huDirectObservationBimolecular2019,toscanoColdControlledChemical2020}.  Laser cooling of molecules is a critical step in realizing many of these applications, which rely on quantum control and coherence, and tremendous advances along these lines have been made in the last decade~\cite{tarbuttLaserCoolingMolecules2018,mccarronLaserCoolingTrapping2018,Isaev2020Review,Fitch2021Review}.  Molecular laser cooling and trapping is now starting to offer full quantum control~\cite{andereggOpticalTweezerArray2019,Cheuk2020Collisions,Burchesky2021Rotational,Anderegg2021Shielding,Holland2022Entanglement,Bao2023Entangle,Anderegg2023CaOHSP}, which has been a powerful driver of atom-based quantum science.

Laser cooling and trapping of atomic and molecular systems relies on the spontaneous optical forces generated through optical cycling -- the continual process of optical excitation of the atom or molecule to an excited electronic state, followed by spontaneous decay. The main challenge in laser cooling of molecules is that an excited electronic state can spontaneously decay to any excited vibrational state in the ground electronic state~\cite{DiRosa2004Molecules}.  In order to continue cycling photons, population in these excited vibrational states must be repumped back to the ground vibrational state, which typically requires 3 to 12 repump lasers~\cite{Fitch2021Review}.  The main approach to addressing this challenge is to work with molecules for which the probability of ``off-diagonal'' decays to excited vibrational states are small~\cite{DiRosa2004Molecules,isaevPolyatomicCandidatesCooling2016,kozyryevProposalLaserCooling2016}.  Nonetheless, leakage to these higher-lying vibrational states is still the main limiting factor in molecular laser cooling and trapping.

Since individual quantum states in the molecule must be addressed for repumping, high-resolution spectroscopy of at least a few lines out of vibrational leakage states must be performed.  A typical approach is to measure the energy of these states at ``low resolution'' ($\sim$100~GHz) via dispersed fluorescence~\cite{mengeshaBranchingRatiosRadiative2020,zhangAccuratePredictionMeasurement2021,Lasner2022SrOH}, whereby the fluorescence of molecular decays from the electronic excited state is dispersed by a diffraction grating and imaged to yield wavelength information.  This method provides a more than sufficient starting point for high-resolution, Doppler-limited spectroscopy ($\sim$10~MHz) needed to identify repumping transitions.

However, a challenge for high-resolution spectroscopy is the low molecular population of high-lying vibrational states in the cryogenic buffer gas environment where these experiments typically start~\cite{hutzlerBufferGasBeam2012}.  The usual solution is to optically pump into vibrational levels of interest  by cycling on laser cooling transitions until population accumulates into unaddressed states, and then performing laser-induced fluorescence spectroscopy.  Though this method is effective, it requires multiple high-power lasers and sufficient interaction time to realize optical cycling. This can be quite challenging for decays at the $\sim 10^{-4} - 10^{-5}$ level, which require cycling $>10^3$ photons to populate the vibrational leakage states. 

In this manuscript, we demonstrate a general method to perform high-resolution spectroscopy in a cryogenic buffer gas cell on excited vibrational states needed for molecular repumping during optical cycling.  We directly increase population in excited vibrational states by implementing optically-driven chemical reactions between metastable atoms and molecules, populating many vibrational states that are not efficiently thermalized by the cryogenic buffer gas~\cite{jadbabaieEnhancedMolecularYield2020}.  We then implement sensitive frequency-modulated (FM) absorption \cite{hallTRANSIENTLASERFREQUENCY2000,bjorklundFrequencymodulationSpectroscopyNew1980,bjorklundFrequencyModulationFM1983b} to perform direct spectroscopy on these states without the need for optical cycling lasers.  We demonstrate this method by measuring the $\tilde{A}^2\Pi_{1/2}(1,0,0) - \tilde{X}^2\Sigma^+(3,0,0)$  transition in $^{174}$YbOH. The notation here is that $\tilde{X}^2\Sigma^+$ is the ground electronic state, $\tilde{A}^2\Pi_{1/2}$ is an excited electronic state, and $(\nu_\mathrm{Yb-O},\nu_\mathrm{Bend},\nu_\mathrm{O-H})$ gives the number of quanta in the Yb-O stretch, bending, and O-H stretch modes, respectively~\cite{kozyryevPrecisionMeasurementTimeReversal2017}. We will use the abbreviated state notation $\tilde{X}(\nu_\mathrm{Yb-O}\nu_\mathrm{Bend}\nu_\mathrm{O-H})$ and $\tilde{A}(\nu_\mathrm{Yb-O}\nu_\mathrm{Bend}\nu_\mathrm{O-H})$ going forward. The $\xe$ state is a laser-cooling leakage channel with branching ratio $\sim7\times10^{-4}$ from the excited $\ag$ state used for cycling and cooling~\cite{zhangAccuratePredictionMeasurement2021}. For comparison, a minimum of four high power lasers would be needed to sufficiently populate the $\xe$ state through optical cycling. 

From the high resolution spectrum we assign 35 transitions and identify the $^PQ_{12}(1)$ and $^PP_{11}(1)$ lines, which can be used to repump losses to the $\xe$ state when laser cooling YbOH. Additionally, we combine the 35 transition frequencies measured here with the 65 previously measured transition frequencies of the $\Ae - \xg$ transition \cite{steimleFieldfreeStarkZeeman2019} and perform a non-linear least squares fit to an effective Hamiltonian model. From the fit, we determine the spectroscopic parameters of the $\xe$  state for the first time and improve the estimate of the spectroscopic parameters of the $\Ae$ state. We achieve high signal-to-noise which could easily be improved further to readily measure other high-lying excited vibrational states for molecular laser cooling in a wide range of species.

\section{\label{sec:Experiment}Experiment}

\begin{figure}[t] 
\includegraphics[width=\columnwidth,keepaspectratio]{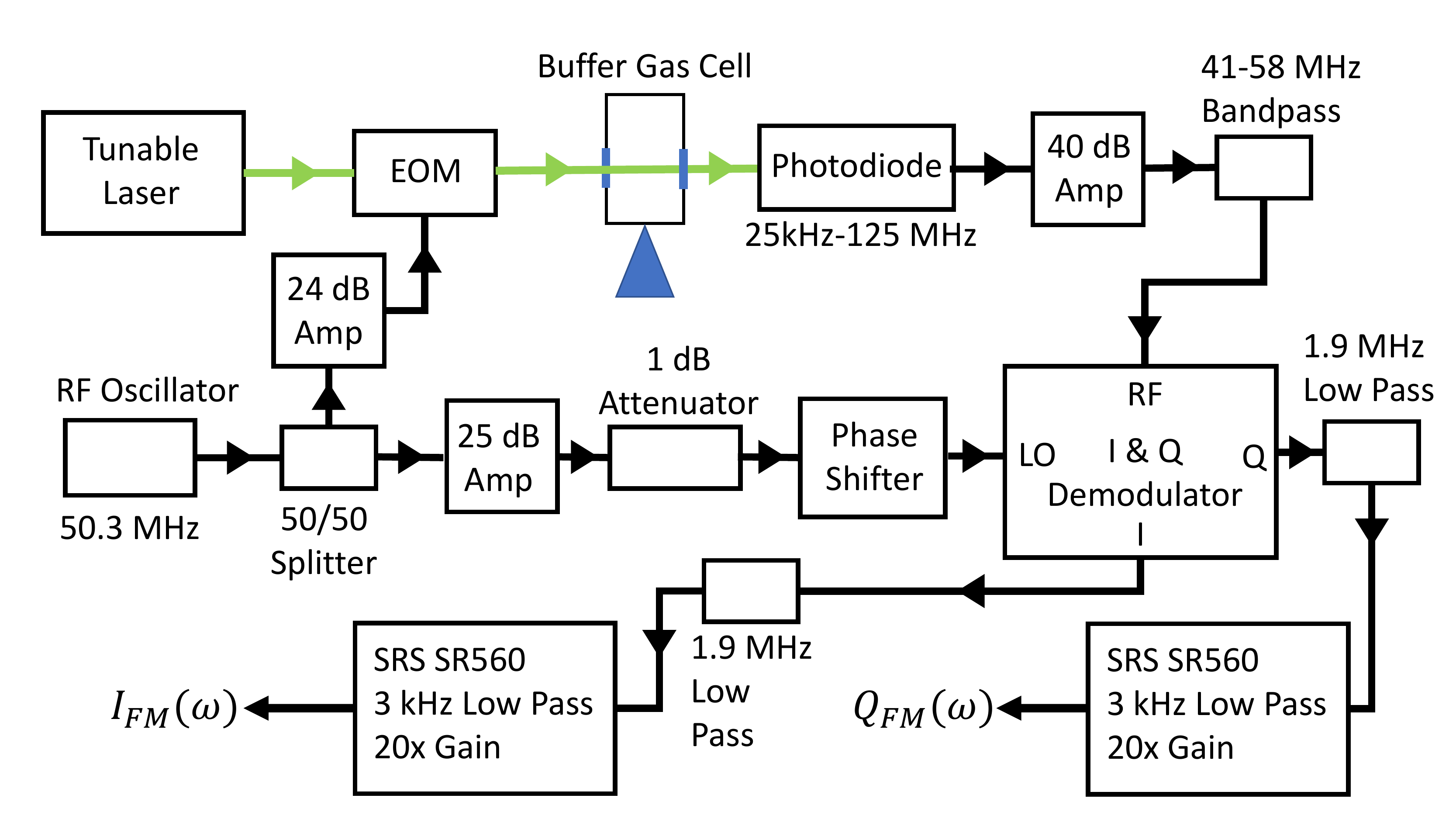}
\caption{\label{fig: FM-setup} Diagram of FM absorption spectroscopy experimental setup. The green lines indicate the laser path and the black lines indicate the rf (or DC after demodulation) signal path. Detailed descriptions of the components and their functions are given in the text.}
\end{figure}

A diagram of the experimental setup is shown in Fig. \ref{fig: FM-setup}. In this work we perform FM spectroscopy inside a cryogenic buffer gas cell~\cite{hutzlerBufferGasBeam2012}. The cryogenic buffer gas source is the same source used in the previous studies of YbOH \cite{jadbabaieEnhancedMolecularYield2020,pilgramFineHyperfineInteractions2021}, and details can be found therein. In short, a copper buffer gas cell with an internal cylindrical bore of 12.7 mm and a length of $\sim$75 mm is cooled to $\sim$4 K. Helium buffer gas is introduced into the cell by a 3.2 mm diameter copper tube and exits the cell from a 5 mm aperture located on the opposite end of the cell.  YbOH molecules are produced by ablating a solid pressed target with a pulsed Nd:YAG laser (532 nm, 42 mJ/pulse). The pressed target is a mixture of Yb and Yb(OH)$_3$ powders as described elsewhere~\cite{jadbabaieEnhancedMolecularYield2020}. YbOH production is increased via laser-enhanced chemical reactions \cite{jadbabaieEnhancedMolecularYield2020}, realized by saturating the $^3P_1\leftarrow{}^1S_0$ $^{174}$Yb transition at 556~nm. Importantly, this process populates high-lying vibrational states which are not efficiently thermalized by the buffer gas~\cite{kozyryevCollisionalRelaxationVibrational2015,jadbabaieEnhancedMolecularYield2020}. The entirety of this work, including molecular production and absorption measurements, occurs inside the buffer gas cell, which has several windows to admit the ablation, chemical production, and absorption lasers.

FM spectroscopy was performed with a 3.5~mW cw laser ($\sim$1 mm in diameter) which was derived via the sum-frequency generation of a cw Ti:Sapph and 1550 nm fiber laser (Sirah Mattise and NKT ADJUSTIK+BOOSTIK combined in a Sirah MixTrain) with a linewidth of $<50$ kHz.  Sidebands are applied to the laser with a resonant electro-optic phase modulator (EOM) (Thorlabs EO-PM-R-50.3-C4). The EOM is driven by a 69.8 mW (18.44 dBm), 50.3 MHz sinusoidal rf drive, corresponding to a modulation depth of $M=0.84$. The EOM was calibrated with a Fabry-P\'{e}rot interferometer to determine the relationship between applied rf power and modulation depth. This calibration was found to be very reproducible.

The laser was tuned to 612 nm to probe the $\Ae\leftarrow\xe$ transition.  This transition was previously observed in fluorescence at low resolution~\cite{mengeshaBranchingRatiosRadiative2020,zhangAccuratePredictionMeasurement2021}, but not at high resolution. The electronic states arise from metal-centered electronic states, very analogous to other laser-coolable diatomics and polyatomics with alkaline-earth (and similar) metals~\cite{isaevPolyatomicCandidatesCooling2016,kozyryevProposalLaserCooling2016}. 

The rf setup for the FM spectroscopy is shown in Fig. \ref{fig: FM-setup}.  A 50.3 MHz rf drive is supplied from a DDS signal generator (Novatech 409B).  The DDS output is split, both driving the EOM via a 24 dB amplifier (Mini-Circuits ZHL-3A+) and acting as a reference for the FM absorption measurement.  The reference signal passes through a 25 dB low-noise amplifier (Mini-Circuits ZX60-P103LN+), a 1 dB attenuator, and voltage controlled phase shifter (Mini-Circuits JSPHS-51+), before entering into the local oscillator (LO) port of the I and Q demodulator (Pulsar Microwave Corp. IDO-03-412). The 25 dB amplifier and 1 dB attenuator set the rf power going into the phase shifter at 7.6 dBm. It was experimentally found that this phase shifter input power optimized the FM signal SNR. Phase shifter input powers higher than $\sim$7 dBm provided no significant improvement in the SNR. The phase shifter allows the phase of the rf reference to be tuned, which allows adjustment of the phase angle, $\theta$, of the demodulated in-phase, $I_{FM}(\omega)$, and in-quadrature, $Q_{FM}(\omega)$, signals. We generally operate at a phase shifter voltage where the $I_{FM}(\omega)$ and $Q_{FM}(\omega)$ signals are approximately equal in magnitude. 

After sidebands are applied to the laser with the EOM, the laser beam passes through the buffer gas cell and is detected with an AC-coupled fast photodiode (New Focus 1801 photoreceiver) with a 25 kHz -- 125 MHz bandwidth. The resulting AC signal is then amplified with a 40 dB low-noise amplifier (Mini-Circuits ZKL-1R5+) and passed through a 41-58 MHz band pass filter before being input into the RF port of the I and Q demodulator. Multiple combinations of a second amplifier and additional bandpass, low-pass, and high-pass filters were tried, and none resulted in improved SNR compared to the single amplifier and bandpass filter. 

The I and Q demodulator essentially consists of two rf mixers and a 90$^\circ$ phase shifter. The output of the I port is the in-phase demodulated DC signal resulting from the mixing of the photodiode signal (RF port) and the rf reference (LO port). The output of the Q port is the in-quadrature demodulated DC signal resulting from the mixing of the photodiode signal (RF port) and the rf reference (LO port) with a 90$^\circ$ phase shift. The outputs of both the I and Q ports are passed through 1.9 MHz low-pass filters and input into SRS SR560 low noise pre-amplifiers. The SR560s are set to have a 12dB/oct 3 kHz low-pass filter and 20x gain. Due to the fact that the molecular pulse is $\sim$ 1 ms long, setting the low-pass filter cutoff any lower begins to filter out the DC FM signal. The outputs of both of the SR560s are the measured providing the $I_{FM}(\omega)$ and $Q_{FM}(\omega)$ signals.

\begin{figure}[ht] 
\includegraphics[width=\columnwidth,keepaspectratio]{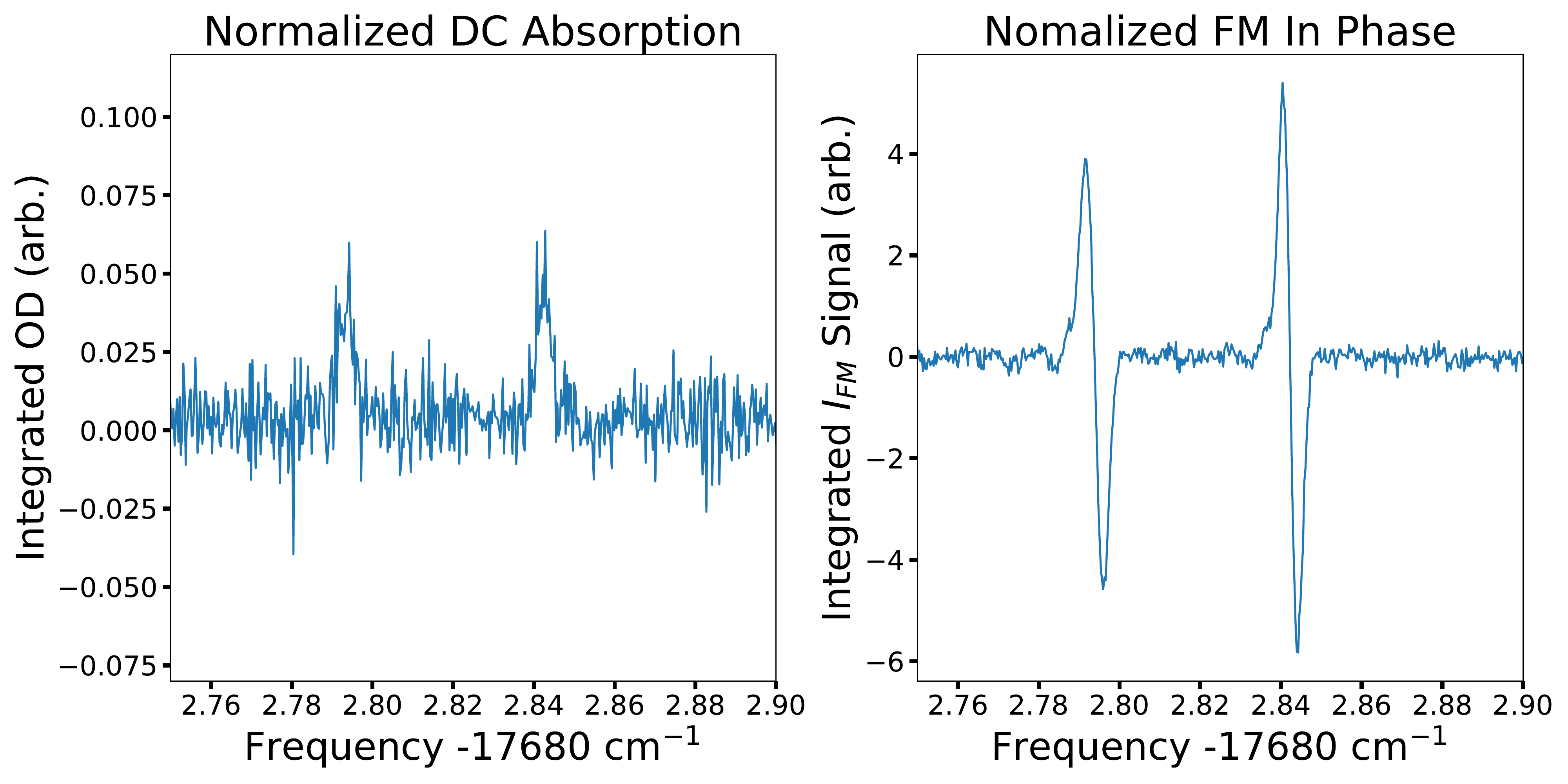}
\caption{\label{fig: FM DC comp slow scan} DC absorption and in-phase FM absorption of two lines of the $[17.68]$ band of YbOH. To record this data, four ablation shots were taken at each frequency and then averaged. A frequency step of 9 MHz was used. Both the integrated DC optical depth and the integrated $I_{FM}$ signal were normalized by the integrated optical depth from a DC absorption probe fixed to the $^RR_{11}(0)$ line of the $\tilde{A}^2\Pi_{1/2}(0,0,0)-\tilde{X}^2\Sigma^+(0,0,0)$ band of YbOH \cite{steimleFieldfreeStarkZeeman2019}.}
\end{figure}

Compared to DC absorption, FM absorption increases the signal-to-noise ratio (SNR) by about an order of magnitude. A comparison of the co-recorded DC and FM absorption signals for two lines of the $[17.68]$ band\footnote{The $[17.68]$ vibronic band of YbOH was identified in Ref. \cite{mengeshaBranchingRatiosRadiative2020} and corresponds to a transition at 17,680 cm$^{-1}$. Assignments of the high resolution spectra have not yet been made.} of YbOH are shown in Fig. \ref{fig: FM DC comp slow scan}. Here, FM absorption provided a factor of $\sim 10$ increase in the SNR. This  improvement in the SNR allows the detection of spectral features that can not be resolved with standard DC absorption techniques.

To record the $\tilde{A}^2\Pi_{1/2}(1,0,0)-\tilde{X}^2\Sigma^+(3,0,0)$ spectrum, the FM spectroscopy laser was continuously scanned and the signal from every five consecutive shots averaged, resulting in $\sim$10 MHz spacing between data points. The frequency of the FM laser is continuously recorded with a HighFinesse wavemeter (WS7-30 VIS/Standard model) which is used to track the relative frequency spacing between data points. The sub-Doppler saturated absorption spectrum of I$_2$ is co-recorded with with light picked off from the spectroscopy laser and used for absolute frequency calibration. Absolute frequency calibration with the sub-Doppler I$_2$ spectrum results in an absolute frequency error of $\lesssim 6$ MHz. 

Shot-to-shot fluctuations in the molecule yield are normalized via a DC absorption measurement using a laser fixed to the to the $^RR_{11}(0)$ line of the $\ag-\xg$ band of $^{174}$YbOH \cite{steimleFieldfreeStarkZeeman2019}. The measured in-phase and in-quadrature FM signals are integrated over the duration of the molecular pulse and scaled by the integrated normalization probe optical depth to produce the in-phase, $I_{FM}(\omega)$, and in-quadrature, $Q_{FM}(\omega)$, spectrum.   

\section{\label{sec:Observation and Assignment}Observation and Assignment}

\begin{figure}[t] 
\includegraphics[width=\columnwidth,keepaspectratio]{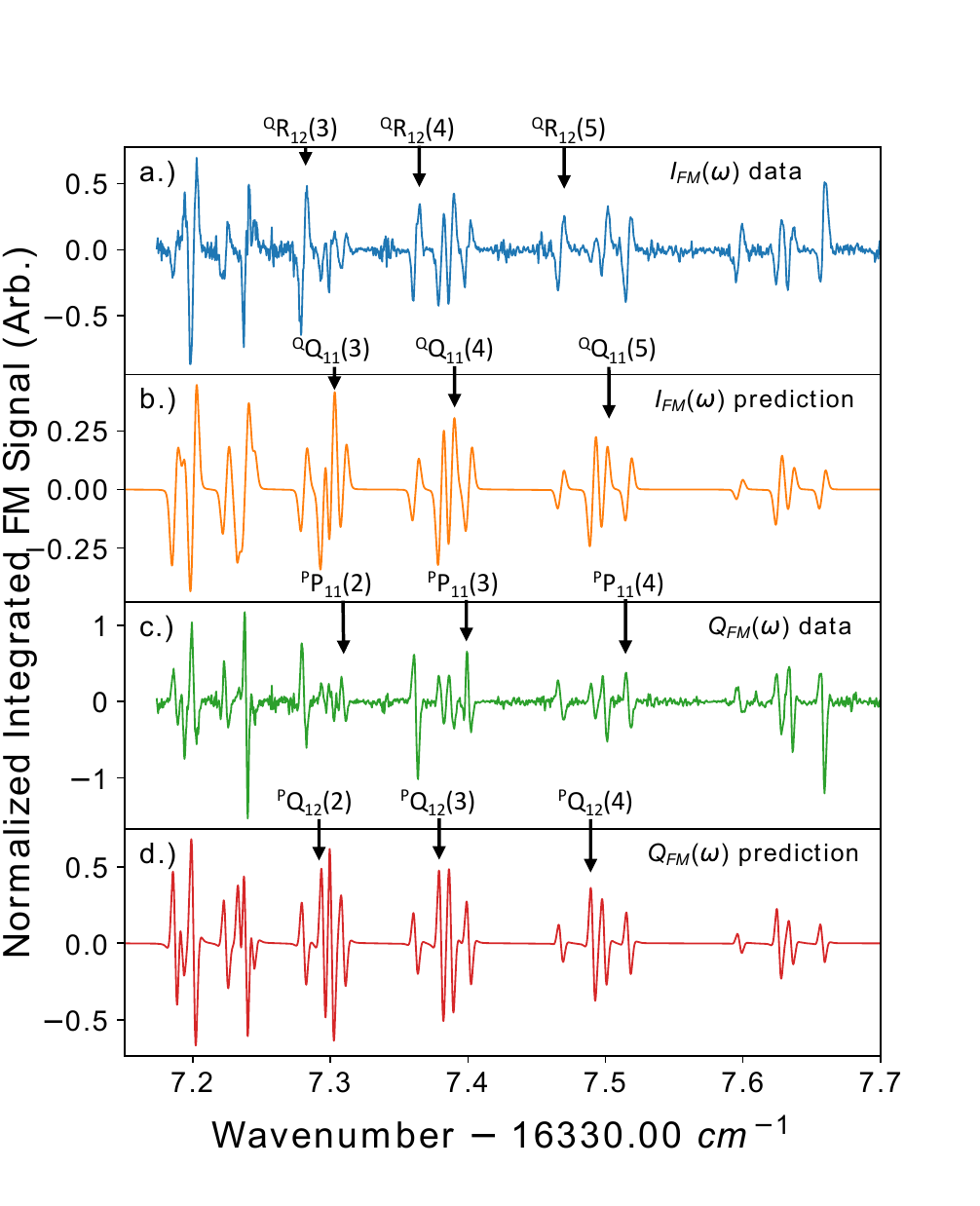}
\caption{\label{fig: X300 A100 bandhead} Measured and predicted FM spectrum in the band head region of the $\Ae-\xe$ band of $^{174}$YbOH. Several features of the $^QR_{12}$, $^QQ_{11}$, $^PP_{11}$, and $^PQ_{12}$ branches are indicated in separate plots for clarity. \textbf{a.)} Measured in-phase FM spectrum, $I_{FM}(\omega)$. \textbf{b.)} Predicted in-phase FM spectrum. \textbf{c.)} Measured in-quadrature FM spectrum, $Q_{FM}(\omega)$. \textbf{d.)} Predicted in-quadrature FM spectrum. The predicted in-phase and in-quadrature FM spectra were obtained using the FM lineshape model described in Appendix \ref{Appx: FM model}. The transition frequencies and relative amplitudes input into the model were calculated using the optimized parameters given in Table \ref{tab: X300 A100 results}. A phase angle of 5.90 radians, FWHM Gaussian absorption linewidth of 108 MHz, and a temperature of 5 K were used in the predictions.}
\end{figure}

The observed in-phase, $I_{FM}(\omega)$, and in-quadrature, $Q_{FM}(\omega)$, FM spectrum of a portion of the band head region of the $\Ae-\xe$ spectrum of $^{174}$YbOH is presented in Fig. \ref{fig: X300 A100 bandhead}. Also presented are the predicted in-phase and in-quadrature FM spectra. The predictions were made using the optimized parameters determined in this study (Table \ref{tab: X300 A100 results}), a phase angle of 5.90 radians, a linewidth of 108 MHz, a temperature of 5 K, and the FM lineshape model described in Appendix \ref{Appx: FM model}. The phase angle and spectral linewidth used in the predictions are the average of the measured phase angles and linewidths determined from a non-linear least squares fit of the FM lineshapes, discussed below.  Note that the SNR of this spectrum is $<1$ if DC absorption is used, or if the chemical enhancement is not used.

The utilization of laser-enhanced chemical reactions allowed the isolation of only the $^{174}$YbOH spectrum. This isolation of the $^{174}$YbOH spectrum is similar to the approach used to isolate the spectra of the odd isotopologues of YbOH in Ref. \cite{pilgramFineHyperfineInteractions2021}. No evidence of H ($\textbf{I}=1/2$)) hyperfine splittings was observed and, therefore, the typical $^2\Pi$ Hund's case (a) - $^2\Sigma^+$ Hund's case (b) branch designation $^{\Delta N}\Delta J_{F_i'F_i''}$ is used to label the measured transitions. For the $\xe$ state, $F_i''=1$ for $J''=N''+1/2$ and $F_i''=2$ for $J''=N''-1/2$. For the $\Ae$ state $F_i'=1$. Here, $N$ is the total non-spin (that is, orbital and rotational) angular momentum, which couples to electron spin $S$ via the spin-rotation interaction to form $J$.

The observed intensities of the $^PQ_{12}$ and $^QQ_{11}$ branch features (Fig. \ref{fig: X300 A100  bandhead}\textbf{a} and \textbf{c}) are weaker than those of the $^QR_{12}$ and $^PP_{11}$ branches. This is in contrast to the predicted intensities (Fig. \ref{fig: X300 A100 bandhead}\textbf{b} and \textbf{d}) where the opposite is the case. This discrepancy between the observed and predicted relative intensities between different branch features was also observed the $\Ae-\xg$ band \cite{steimleFieldfreeStarkZeeman2019}. The reduction of the intensities of the $^PQ_{12}$ and $^QQ_{11}$ branch features (or the increase in the intensity of the  $^QR_{12}$ and $^PP_{11}$ branch features) may be due to perturbations arising from the mixing of the $\tilde{X}^2\Sigma^+$ and/or the $\tilde{A}^2\Pi_{1/2}$ states with other vibronic states.

\begin{figure}[t] 
\includegraphics[width=\columnwidth,keepaspectratio]{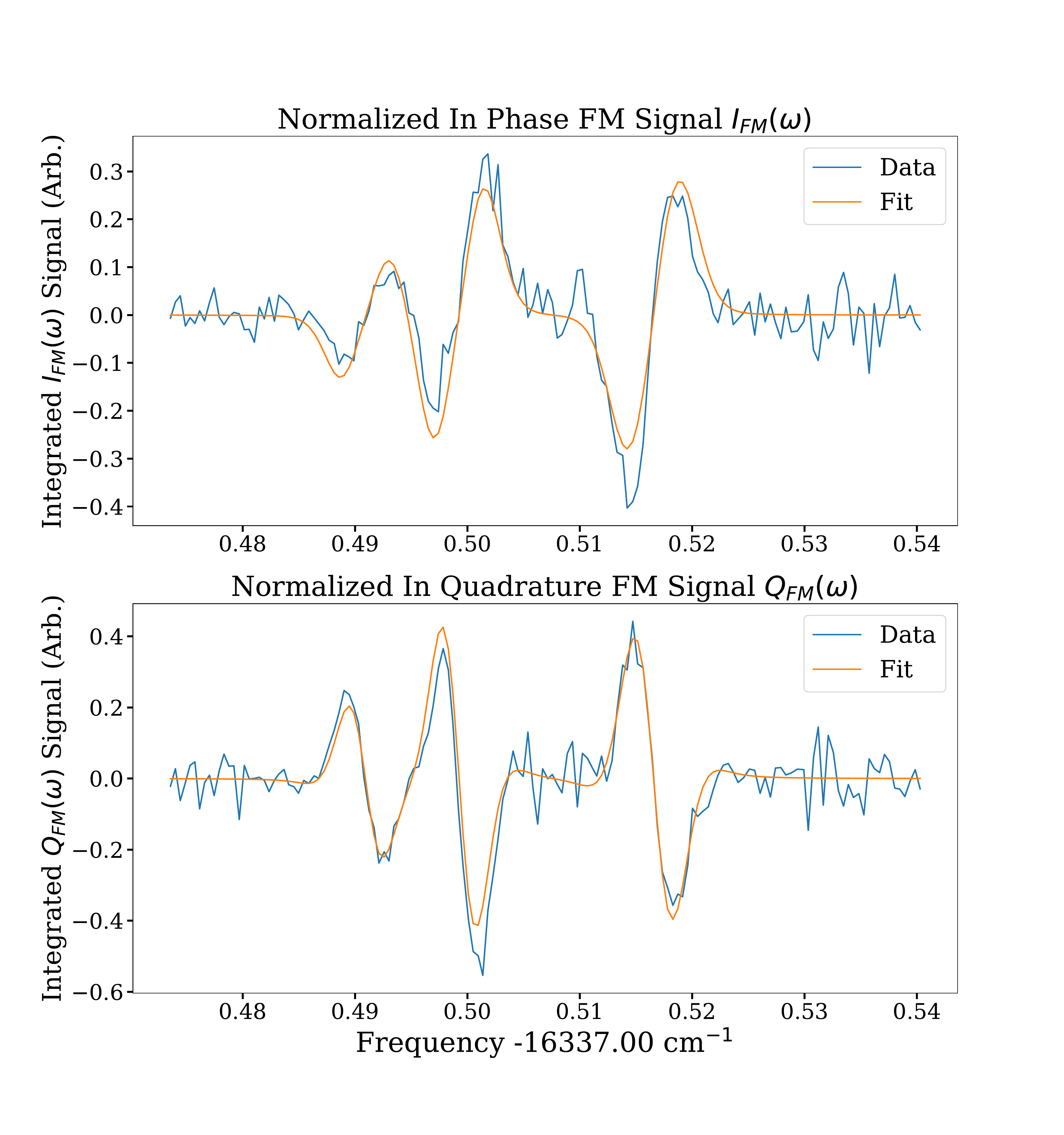}
\caption{\label{fig: X300 A100 fit example} Simultaneous fit of the in-phase and in-quadrature FM spectrum of the (in order of increasing frequency) $^PQ_{12}(4)$, $^QQ_{11}(5)$, and $^PP_{11}(4)$ lines of the $\Ae-\xe$ band of $^{174}$YbOH to the FM lineshape model given in Appendix \ref{Appx: FM model}. The line center, linewidth, and relative amplitude of the Gaussian absorption lineshape of each transition, as well as the overall phase angle between the in-phase and in-quadrature signals were floated in the fit. The line centers resulting from the fit provide a measurement of each transition frequency. }
\end{figure}

The transition wavenumbers were determined via a simultaneous non-linear least squares fit of the measured in-phase and in-quadrature FM lineshapes to the lineshape model described in Appendix \ref{Appx: FM model}. The simultaneous fit of the $^PQ_{12}(4)$, $^QQ_{11}(5)$, and $^PP_{11}(4)$ lines is presented in Fig. \ref{fig: X300 A100 fit example}. The line centers (transition wavenumber), linewidths (FWHM), and relative heights of the Gaussian absorption profiles, as well as the phase angle between the in-phase and in-quadrature FM signals, were floated in the fit. The data set was cut so that a minimum number of spectral features were fit at a single time. An average phase angle of 5.90 radians and an average linewidth of 108 MHz were measured from the fits of the FM data. This linewidth is consistent with the previously measured DC absorption linewidth of $\sim90$ MHz.

Assignments of the spectral features were made using both combination differences and spectral predictions. Combination differences using $\Ae$ energy levels were used to assign low $J$ spectral features and determine the energies of the $N=1$, $J=3/2$, $N=2$, $J=3/2$ and $N=2$, $J=5/2$, levels of the $\xe$ state.  These energy levels were then used to estimate the origin, $T_0''$, the rotational constant, $B''$, and spin rotation parameter, $\gamma ''$, of the $\xe$ state. These estimated parameters were then used in conjunction with the previously determined parameters for the $\Ae$ state \cite{steimleFieldfreeStarkZeeman2019} to predict the $\Ae-\xe$ spectrum, with which the remaining spectral assignments were made. The 35 measured transitions wavenumbers, along with the assignments and associated quantum numbers, are presented in Table \ref{tab: X300 - A100 FM lines}. Also presented in Table \ref{tab: X300 - A100 FM lines} are the differences between the observed and calculated transition frequencies. The calculated transition frequencies were obtained using the optimized parameters given in Table \ref{tab: X300 A100 results}. In addition to the 35 assigned transitions, 3 unassigned transitions were also observed and are listed in Table \ref{tab: X300 - A100 FM lines}.

\section{\label{sec:Analysis}Analysis}

The 35 transition frequencies of the $\Ae-\xe$ band, measured here, were combined with the 65 previously measured transitions frequencies of the $\Ae-\xg$ band (measured via molecular beam LIF) \cite{steimleFieldfreeStarkZeeman2019} and used as inputs in a non-linear least squares fitting procedure. The two data sets share the same excited $\Ae$ state and simultaneous fit of both data sets was performed in order to obtain the optimized parameters of the $\xe$ state, as well as improved parameters for the $\Ae$ state. The $\Ae-\xe$ FM transition frequencies were determined to approximately a factor of two higher precision as compared to the molecular beam LIF data of the $\Ae-\xg$ band~\cite{steimleFieldfreeStarkZeeman2019} and, therefore, the FM data of the $\Ae-\xe$ band was weighted twice that of the previously recorded $\Ae-\xg$ molecular beam LIF data in the fitting procedure. The increased precision of the FM data is due to both the high intrinsic sensitivity of the FM absorption method for measuring transition frequencies (measuring simultaneous zero crossing of in-phase and in-quadrature line shapes) as well as the isolation of the $^{174}$YbOH spectrum from that of the the other isotopologues with the laser-enhanced chemical reactions. 

The energy levels of the $\xg$, $\xe$, and the $\Ae$ states were modeled using an effective Hamiltonian approach~\cite{brownRotationalSpectroscopyDiatomic2003}. The effective Hamiltonian used to model the $\xg$ state is 
\begin{equation}
\begin{split}
\hat{H}_\mathrm{eff}^{[\xg]} = &B\textbf{R}^2 - D \textbf{R}^2\textbf{R}^2 +\gamma \textbf{N}\cdot \textbf{S} \\
&+ \gamma_D\left[ \textbf{N}\cdot \textbf{S},\textbf{R}^2\right]_+,
\end{split}
\end{equation}
 which accounts for rotation ($B$), centrifugal distortions ($D$), spin-rotation ($\gamma$), and spin-rotation centrifugal distortions ($\gamma_D$).  Here, \textbf{R} is the end-over-end rotation of the molecule, \textbf{N} is the total non-spin angular momentum, \textbf{S} is the electron spin, and $[\;]_+$ is an anti-commutator. In the least-squares fit and for all spectral predictions, the parameters of the $\xg$ state were fixed to the values determined with PPMODR microwave spectroscopy \cite{nakhatePureRotationalSpectrum2019}. The effective Hamiltonian used to model the $\xe$ state is
\begin{equation}
\hat{H}_\mathrm{eff}^{[\xe]} = T_0 + B\textbf{R}^2 - D\textbf{R}^2\textbf{R}^2 + \gamma\textbf{N}\cdot\textbf{S}.
\end{equation}This effective Hamiltonian is the same as that of the  $\xg$ state, with the addition of an origin ($T_0$) to account for the vibrational energy, and the removal of the term accounting for the spin-rotation centrifugal distortions. The FM spectrum recorded in the cryogenic buffer gas cell only probes lower $J$ transitions, and is not sensitive to spin-rotation centrifugal distortions. The effective Hamiltonian used to model the $\Ae$ state is
\begin{equation}
\begin{split}
\hat{H}&_\mathrm{eff}^{[\Ae]} = \\
&T_0 + AL_zS_z+ B\textbf{R}^2 -D\textbf{R}^2\textbf{R}^2 \\
&+ \frac{1}{2}(p+2q)(J_+S_+e^{-2i\theta} + J_-S_-e^{+2i\theta}) \\
&+ (p+2q)_D\left[\frac{1}{2}(J_+S_+e^{-2i\theta} + J_-S_-e^{+2i\theta}), \textbf{R}^2 \right]_+,
\end{split}
\end{equation}which accounts for the origin of the electronic state ($T_0$), spin-orbit ($A$), rotation ($B$), centrifugal distortions ($D$), $\Lambda$-doubling ($p+2q$), and $\Lambda$-doubling centrifugal distortions ($(p+2q)_D$). Since no evidence of H hyperfine splittings were observed, hyperfine interactions were not included in the effective Hamiltonians.

Though the $\xg$ and $\xe$ states are best described by a Hund's case (b) basis, for computational convenience all effective Hamiltonians were constructed in a Hund's case (a) basis, $|\eta, \Lambda\rangle|S, \Sigma\rangle|J,\Omega\rangle$. The energy levels and eigenstates of the $\xg$ and $\xe$ states were determined by constructing and diagonalizing the full 34$\times$34 ($2(N_{max}+1)$) Hamiltonian for all $N=0$ to $N=16$ rotational levels while the energy levels and eigenstates of the $\Ae$ state were determined by construction and diagonalizing the full 66$\times$66 ($2(2N_{max}+1$) Hamiltonian for all $N=1$ to $N=16$ rotational levels. The matrix elements used in the calculation of the effective Hamiltonians were taken from Ref. \cite{brownRotationalSpectroscopyDiatomic2003,brownAnalysisHyperfineInteractions1978}.

The origin ($T_0''$), rotational parameter ($B''$), and the spin-rotation parameter ($\gamma''$) of the $\xe$ state and the origin ($T_0'$), rotational parameter ($B'$), $\Lambda$-doubling parameter $(p+2q)'$, and $\Lambda$-doubling centrifugal distortion parameter ($(p+2q)'_D$) of the $\tilde{
A}^2\Pi_{1/2}(1,0,0)$ state were floated (a total of seven parameters) in the final least-squares fit to the measured transition frequencies. The spin-orbit parameter, $A$, of the $\Ae$ state, was fixed to the value from the high temperature analysis \cite{melvilleVisibleLaserExcitation2001}. Fits floating various parameters were performed and an f-test with a 95\% confidence interval was used to determine if floating additional parameters (such as $(p+2q)'_D$, $D''$, or $D'$) was statistically justified. While the f-test indicated floating $D$ in both the $\xe$ and $\Ae$ states was statistically justified, the error in the resulting fit $D$ parameters was $\sim20\%$ of the fitted value. This indicated that floating the $D$ parameters resulted in values for $D''$ and $D'$ that were not well determined. Therefore, the value of $D$ in the $\xe$ state was fixed to the the value extrapolated from the $D$ values of the $\xg$ and $\tilde{X}(100)$ states using the expected vibrational dependence \cite{bernathSpectraAtomsMolecules2005}. The value of $D$ in the $\Ae$ state was fixed to the extrapolated value given in Ref. \cite{steimleFieldfreeStarkZeeman2019}. 

The optimized parameters of the $\xe$ and $\Ae$ states resulting from the least-squares fit to the transition frequencies are presented in Table \ref{tab: X300 A100 results}. Also, presented in Table \ref{tab: X300 A100 results} are the parameters of the $\tilde{X}(100)$ state \cite{steimleFieldfreeStarkZeeman2019} for comparison. The fit resulted in an RMS of the residuals of 25 MHz (0.00084 cm$^{-1}$), which is commensurate with the measurement uncertainty of the combined data set. The difference between the observed and calculated transition frequencies (fit residuals) for the $\Ae-\xe$ and $\Ae-\xg$ bands are given in Table \ref{tab: X300 - A100 FM lines} and \ref{tab: X000 - A100 lines}, respectively.

\begin{table*}[ht]
\caption{\label{tab: X300 A100 results} Spectroscopic parameters of the $\xg$, $\xe$, and $\Ae$ states of $^{174}$YbOH.  The parameters of the $\tilde{X}(100)$ state are also presented for comparison. All values are in cm$^{-1}$. Values in parenthesis are the standard errors resulting from the combined fit of the $\Ae-\xg$ and  $\Ae-\xe$ bands.}
\begin{ruledtabular}
\begin{tabular}{c c c c c c c c}
 & &\multicolumn{3}{c}{Vibrational State}\\ \cline{3-5}\\
  &  Parameter & (0,0,0) &(1,0,0) & (3,0,0) & & Parameter & (1,0,0)\\ 
\hline
 &  $T_0''$ & -- & 529.3269\footnotemark[2] & 1570.6697(2) & & $T_0'$ & 18582.8707(1)  \\
$\mathbf{\tilde{X}^2\Sigma^+}$ & $B''$ & 0.245116257\footnotemark[1] & 0.243681\footnotemark[2] & 0.240795(4) & $\mathbf{\tilde{A}^2\Pi_{1/2}}$  & $A'$ &  1350\footnotemark[4] (fixed)\\
&$10^7D''$ & 2.029\footnotemark[1] & 2.168\footnotemark[2] & 2.45\footnotemark[3] (fixed) & & $B'$ & 0.253197(2) \\
& $\gamma ''$ & --0.002707\footnotemark[1] & --0.00369\footnotemark[2] & --0.00575(3) &  &$10^7D'$ & 2.478\footnotemark[2] (fixed)\\
& $10^7\gamma ''$ & 1.59\footnotemark[1] & -- & -- & & $(p+2q)'$ & --0.53459(4) \\
&  & & & & & $10^6(p+2q)'_D$ & --17.3(3) \\
\end{tabular}
\end{ruledtabular}
\\
\footnotetext[1]{Fixed to PPMODR values in fit, Ref \cite{nakhatePureRotationalSpectrum2019}}.
\footnotetext[2]{From Ref. \cite{steimleFieldfreeStarkZeeman2019}}
\footnotetext[3]{Fixed to value extrapolated from that of the $\xg$ and $(100)$ states}
\footnotetext[4]{Fixed to value from high temperature analysis \cite{melvilleVisibleLaserExcitation2001}}
\end{table*}

Spectral predictions were made in the following manner: First, the transition dipole moment matrix is calculated in a Hund's case (a) basis and cross multiplied by the eigenvectors to determine the transition dipole moments. The relative transition amplitudes are given by the product of the square of the transition moment and a Boltzmann factor. To produce simulated FM spectra, such as those shown in Fig. \ref{fig: X300 A100 bandhead}, the total absorption lineshape, $\delta_{tot}(\omega)$, is calculated by summing individual Gaussian lineshapes for each transition, using Eq. \ref{eq: absorption total}. The transition frequencies and relative amplitudes from the spectral predictions are used for the line centers and amplitudes of the Gaussian lineshapes, and a FWHM of 108 MHz (the averaged measured linewidth) is used as the linewidth for all transitions. The total dispersion lineshape, $\phi_{tot}(\omega)$ is calculated in a similar manner using Eq. \ref{eq: dispersion total}. Finally, the simulated in-phase, $I_{FM}(\omega)$, and in-quadrature, $Q_{FM}(\omega)$, FM lineshapes are calculated using Eq. \ref{eq: A FM 2nd} and \ref{eq: D FM 2nd} and a phase angle of 5.90 radians, the average of the measured phase angle.

\section{\label{sec:Discussion}Discussion}

The main goals of this study were to demonstrate the utility of laser-enhancement and FM spectroscopy for measuring transitions originating from excited vibrational states, to identify the $\xe$ repumping transitions needed to laser cool and trap YbOH, and to determine the spectroscopic parameters of the $\xe$ state. The measured FM spectrum of the $\Ae-\xe$ band demonstrates that the chemical enhancement provides enough molecular population to observe transitions originating from excited vibrational states. Additionally, the in-buffer-gas-cell FM absorption technique provides the needed sensitivity to observe the weak $\Ae-\xe$ band. This combination of chemical enhancement and in-cell FM absorption spectroscopy is a promising technique for measuring transitions originating from excited vibrational states, in both YbOH and in other molecules. 

\begin{figure}[ht] 
\includegraphics[width=\columnwidth,keepaspectratio]{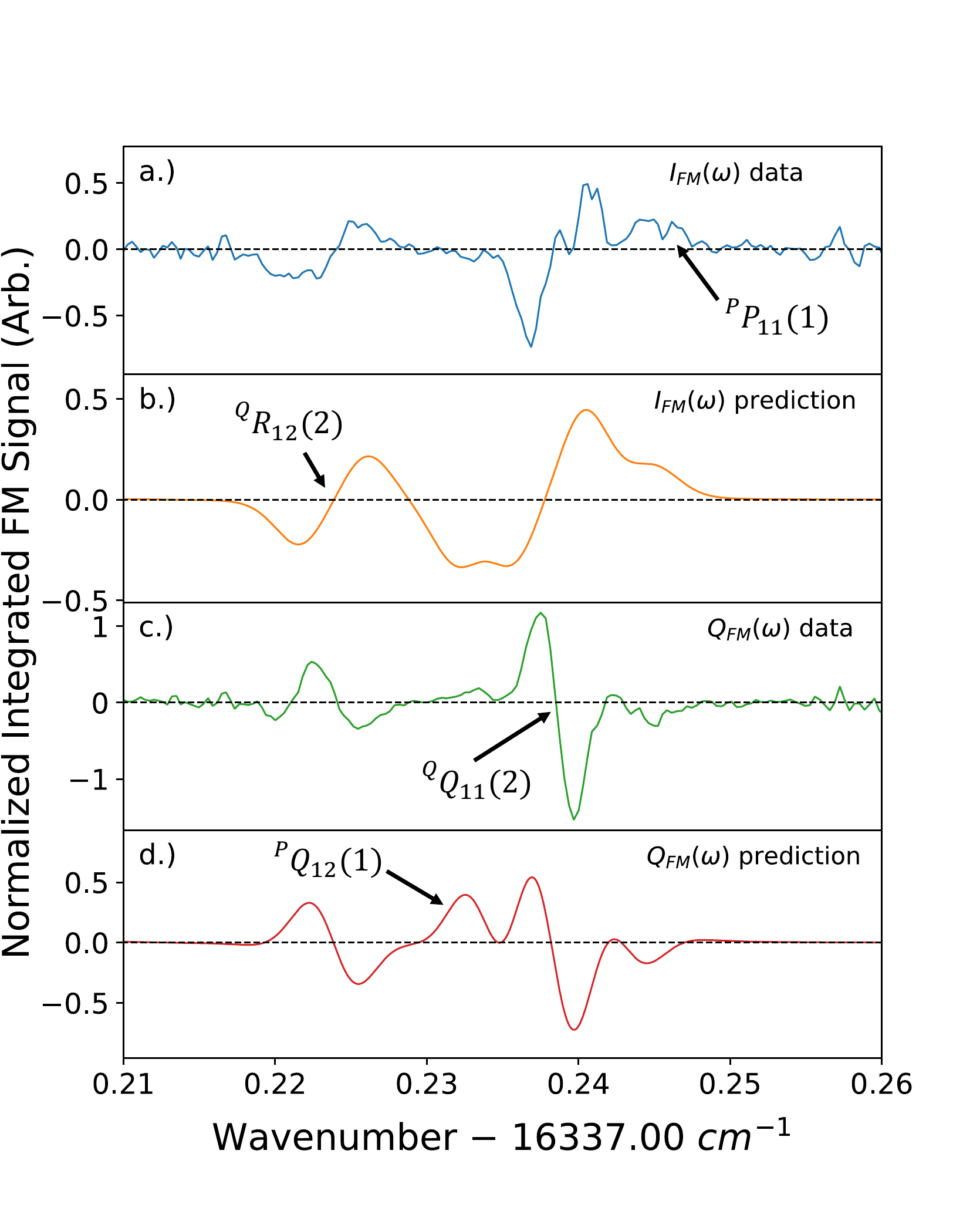}
\caption{\label{fig: X300 A100 Q12(1) line} Measured and predicted FM spectrum of the $^QR_{12}(2)$, $^PQ_{12}(1)$, $^QQ_{11}(2)$ and $^PP_{11}(1)$ transitions of the $\Ae-\xe$ band of $^{174}$YbOH.  Some of the lines appear to be ``missing'' as discussed in the text. \textbf{a.)} The measured in-phase FM spectrum, $I_{FM}(\omega)$. \textbf{b.)} The predicted in-phase FM spectrum. \textbf{c.)} The measured in-quadrature FM spectrum, $Q_{FM}(\omega)$. \textbf{d.)} The predicted in-quadrature FM spectrum. Predictions were made using optimized parameters from Table \ref{tab: X300 A100 results}, a Gaussian FWHM linewidth of 108 MHz, a phase angle of 5.90 radians, and a temperature of 5 K. Each transition is indicated in a different plot for clarity.} 
\end{figure}

The primary interest in the $\Ae-\xe$ band of YbOH stems from its utility as a repumping pathway to return population lost to the $\xe$ state back into the cooling cycle. Repumping is accomplished by driving the rotationally closed $^PQ_{12}(1)$ and $^PP_{11}(1)$ transitions simultaneously \cite{stuhlMagnetoopticalTrapPolar2008}. While the data analysis assigned the $^PP_{11}(1)$ line to the transition at 16337.2431 cm$^{-1}$, no assignment for the $^PQ_{12}(1)$ line was originally made. The prediction of the $\Ae-\xe$ spectrum using the optimized parameters from Table \ref{tab: X300 A100 results} places the $^PQ_{12}(1)$ line at 16337.2342 cm$^{-1}$. The in-phase and in-quadrature FM data and simulated spectra, in the region of the predicted $^PQ_{12}(1)$ line and the measured $^QR_{12}(2)$,$^QQ_{11}(2)$ and $^PP_{11}(1)$ lines, are presented in Fig. \ref{fig: X300 A100 Q12(1) line}. The simulated FM spectra were calculated using the optimized  parameters, a Gaussian FWHM linewidth of 108 MHz, a phase angle of 5.90 radians, and a temperature of 5 K. 

When comparing the prediction to the data, particularly the in-phase data, the $^PQ_{12}(1)$ line appears to be missing. However, upon closer examination, the asymmetry in the in-phase and in-quadrature peak heights of the $^QQ_{11}(2)$ line, as well as the small positive peak to the red of the $^QQ_{11}(2)$ line in the in-quadrature data, indicate that there is a small spectral feature just to the red of, and blended with, the $^QQ_{11}(2)$ feature. When fitting the data in this region to the FM lineshape model, fits both with and without a line to account for this small feature were performed (fits to 3 or 4 spectral features). Adding a line to account for this small feature did not improve the fit. Therefore, a line accounting for this small blended feature was not added to the fit and no transition wavenumber was assigned to this feature. When examining the FM data and the prediction made with the optimized parameters, particularly the in-quadrature data and prediction, this small feature is almost directly overlapped with the predicted $^PQ_{12}(1)$ line. Therefore, even though the observed intensity is much smaller than the predicted intensity, this small feature can be assigned to the $^PQ_{12}(1)$ transition.

The blending of this $^PQ_{12}(1)$ feature with the much more intense $^QQ_{11}(2)$ line precludes measurement of the transition frequency via fitting to the FM lineshape or by zero-crossing measurement. Instead the transition frequency was measured by noting that the maximum of the positive peak of the in-quadrature signal is $\sim$ 50.3 MHz to the red of the center of the Gaussian absorption lineshape. Using this method the $^PQ_{12}(1)$ line was assigned to $16337.2348 \pm 0.0036$ cm$^{-1}$, which disagrees with the predicted value by 18 MHz (0.0006 cm$^{-1}$). The generous error of one average measured linewidth is given due to the approximate method used for measuring the transition frequency. The assigned valued of the $^PQ_{12}(1)$ and $^PP_{11}(1)$ transitions can be immediately used for laser cooling YbOH.

The combined fit of the $\Ae-\xe$ and $\Ae-\xg$ data sets allowed the parameters of the $\xe$ state to be determined for the first time. The determined value of the origin, $T_0 = 1570.6697(2)$ cm$^{-1}$, matches the value of $1572(5)$ cm$^{-1}$ measured using a dispersed LIF technique \cite{zhangAccuratePredictionMeasurement}. The vibrational dependence of the rotational constant, $B$, is \cite{bernathSpectraAtomsMolecules2005}
\begin{equation}\label{eq: B vib dependance}
B_{\nu} = B_e-\alpha_e(\nu+1/2).
\end{equation}
Fitting the measured values of $B$ for the $\xg$ and $\tilde{X}(100)$ states to Eq. \ref{eq: B vib dependance} results in $B_e = 0.245834$ cm$^{-1}$ and $\alpha_e = 0.001435$ cm$^{-1}$. Using these values in Eq. \ref{eq: B vib dependance} predicts $B$ for the $\xe$ state of $0.240810$ cm$^{-1}$. This is in fairly good agreement with the measured value of $0.240795(4)$ cm$^{-1}$, especially considering the quadratic and higher order terms in Eq. \ref{eq: B vib dependance} were ignored. 

The value of the spin-rotation parameter in the $\xe$ state of YbOH, determined here, is negative, as is the case in the $\xg$, $\tilde{X}(100)$, and $\tilde{X}(010)$ states \cite{nakhatePureRotationalSpectrum2019,steimleFieldfreeStarkZeeman2019,jadbabaieCharacterizingFundamentalBending2023}. These negative spin-rotation parameters are the same sign, though larger in magnitude, than the measured spin-rotation parameters of the $X^2\Sigma^+$ state of the isoelectronic molecule YbF \cite{lim557X2S5612017}. In the $\tilde{X}^2\Sigma^+$ state, the value of $\gamma$ is dominated by second order contributions which result from the mixing of the $\tilde{X}^2\Sigma^+$ state with excited $^2\Pi_{1/2}$ states through the combination of the rotational and spin orbit interactions. The second order contributions to $\gamma$ are given by \cite{brownRotationalSpectroscopyDiatomic2003,lim557X2S5612017}
\begin{multline} \label{eq: gamma Sigma state2}
\gamma^{(2)} = 2 \sum_{^2\Pi,\,\nu'} |\langle\nu''|\nu'\rangle|^2\times \\ \frac{\langle ^2\Sigma^+_{-1/2}|BL_-|^2\Pi_{1/2}\rangle\langle^2\Pi_{1/2}|\sum_i a_il_i^+s_i^-|^2\Sigma^+_{1/2}\rangle}{E_{\Pi,\,\nu'}-E_{\Sigma,\,\nu''}},
\end{multline}
where the sum is taken over all excited $^2\Pi$ states and all vibrational levels, $\nu'$, of each excited $^2\Pi$ state. Here, $\nu''$ is the vibrational level of the $^2\Sigma^+$ state and $|\langle\nu''|\nu'\rangle|^2$ is the Franck-Condon factor (FCF) between the $\nu''$ and $\nu'$ levels of the $^2\Sigma^+$ and $^2\Pi$ states. In Eq. \ref{eq: gamma Sigma state2} $L_-$ is the total electron angular momentum lowering operator, $l_i^+$ and $s_i^-$ are the raising and lowering operators for the single electron angular momentum and spin respectively, and the sum is taken over all electrons. 

In the alkaline earth mono-halides (e.g., CaF) the observed spin-rotation parameter, $\gamma$, of the $X^2\Sigma^+$ state can be quantitatively predicted using Eq. \ref{eq: gamma Sigma state2} and the known electronic state distribution. Specifically, $\gamma(X^2\Sigma^+)$ for CaF (=0.0131 cm$^{-1}$) is readily predicted by including only the $X^2\Sigma^+\leftrightarrow A^2\Pi_{1/2}$ interactions in Eq. \ref{eq: gamma Sigma state2} \cite{domailleMicrowaveOpticalDouble}. Like YbOH, the $X^2\Sigma^+$ and $A^2\Pi_{1/2}$ states of the alkaline earth mono-halides are atomic in nature and can be well approximated by linear combination of atomic orbitals. Therefore, if the electronic state distributions and FCFs for YbOH are reasonably well known, $\gamma$ can be estimated using Eq. \ref{eq: gamma Sigma state2} and the known atomic parameters of the Yb$^+$ ion.

When comparing the determined spin-rotation parameter of YbOH with that of the
alkaline earth mono-halides, the most striking difference is that it is of opposite sign, being
negative as opposed to positive. In Ref. \cite{nakhatePureRotationalSpectrum2019}, the negative spin-rotation parameter in the $\tilde{X}^2\Sigma^+$ state of YbOH was attributed to contributions from electronic states with a leading electron configuration which has a hole in the $4f$ core. If we assume that only the $\tilde{
A}^2\Pi_{1/2}$ state and states with an [Xe]$4f^{13}\sigma^2_{Yb^+(6s6p)}$ electronic configuration contribute to the value of $\gamma$, we can estimate the contributions from these $4f^{13}$ states, and their vibrational dependence, by first estimating the contributions from the $\tilde{A}^2\Pi_{1/2}$ state.

Due to the atomic nature of the electronic states of YbOH, the wavefunction of the $\tilde{X}^2\Sigma^+$ state can then be approximated by
\begin{equation}\label{eq: X state wavefunction2}
|\tilde{X}^2\Sigma^+\rangle\approx x_s|6s\sigma\rangle + x_p|6p\sigma\rangle,
\end{equation}
where $|6s\sigma\rangle$ and $|6p\sigma\rangle$ are the Yb$^+$ $6s\sigma$ and $6p\pi$ atomic orbitals. Measurements of the $b_F$ hyperfine parameter of the odd isotopologues \cite{pilgramFineHyperfineInteractions2021} indicate that $|x_s|^2\approx0.54$ and assuming contributions from other atomic orbitals (e.g. $5d\sigma$) are small, $<1\%$, gives $|x_p|^2\approx0.46$. 
The wavefunction of the $\tilde{A}^2\Pi_{1/2}$ state can be approximated by
\begin{equation}\label{eq:A state wavefunction2}
|\tilde{A}^2\Pi_{1/2}\rangle\approx a_p|6p\pi\rangle + a_d|5d\pi\rangle,
\end{equation}where $|6p\pi\rangle$ and $|5d\pi\rangle$ are the Yb$^+$ atomic $6p\pi$ and $5d\pi$ orbitals.  If we parameterize the atomic spin orbit interaction as $\zeta_{n,l}l\cdot s$ \cite{sauerAnomalousSpinRotationCoupling1995} and use the Yb$^+$ atomic ion values, $\zeta_{6p}=2220$ cm$^{-1}$ and $\zeta_{5d}=549$ cm$^{-1}$ \cite{barrowAnalysisOpticalSpectrum1975}, as well as the measured spin orbit parameter of the $\tilde{A}^2\Pi$ state of YbOH, $A=1350$ cm$^{-1}$ \cite{melvilleVisibleLaserExcitation2001}, we estimate $|a_p|^2\approx 0.48$ and $|a_d|^2\approx 0.52$. Using these wavefunctions, the electronic matrix elements in Eq. \ref{eq: gamma Sigma state2} are
\begin{equation}
\begin{split}
&\langle\tilde{X}^2\Sigma^+|BL_-|\tilde{A}^2\Pi_{1/2}\rangle\langle\tilde{A}^2\Pi_{1/2}|\Sigma_i a_il_i^+s_i^-|\tilde{X}^2\Sigma^+\rangle \\
&\quad= B|x_p|^2|a_p|^2\langle6p\sigma|l_i^-|6p\pi\rangle\langle6p\pi|\zeta_{6p}l_i^+|6p\sigma\rangle\\
&\quad=2B|x_p|^2|a_p|^2\zeta_{6p},
\end{split}
\end{equation}
where we have used the pure precession hypothesis \cite{brownRotationalSpectroscopyDiatomic2003} to evaluate the atomic matrix elements. 

The FCFs, $|\langle\nu''|\nu'\rangle|^2$, can be reliably estimated using the measured value of the stretching vibrational frequency ($\omega_{\nu_1}=529.3269$ cm$^{-1}$), the measured bond lengths ($r_e(\xg)=2.0397 \:\mathring{A}$ and $r_e(\ag)=2.0062 \:\mathring{A}$) and the harmonic approximation. The bending and O-H stretching modes can be neglected as the FCFs between them and the stretching (or (0,0,0)) states are negligible. In the harmonic approximation, the energies of the $\tilde{A}^2\Pi_{1/2}$ state are $E_{\Pi,\,\nu'} = T_0(\tilde{A}^2\Pi_{1/2}(0,0,0))+\omega_{\nu_1}\nu_1$. Using the values of $|x_p|^2$, $|a_p|^2$, $\zeta_{6p}$, and $\omega_{\nu_1}$ indicated above, the measured value of  $T_0(\ag)$ \cite{steimleFieldfreeStarkZeeman2019}, the values of $B$ and $T_0=E_{\Sigma,\,\nu''}$ given in Table \ref{tab: X300 A100 results}, and the FCFs calculated in the harmonic approximation with Eq. \ref{eq: gamma Sigma state2}, gives $\gamma_{\nu_1=0}(\tilde{A}^2\Pi_{1/2})=0.0276$ cm$^{-1}$, $\gamma_{\nu_1=1}(\tilde{A}^2\Pi_{1/2})=0.02737$ cm$^{-1}$ and $\gamma_{\nu_1=3}(\tilde{A}^2\Pi_{1/2})=0.02686$ cm$^{-1}$. 

These positive contributions from the $\tilde{A}^2\Pi_{1/2}$ state cannot account for the observed negative values of $\gamma$, nor would adding in additional $^2\Pi$ states, as those contributions would be positive as well. As was discussed in Ref. \cite{nakhatePureRotationalSpectrum2019}, the negative values of $\gamma$ in the $\tilde{X}^2\Sigma^+$ state are the result of perturbing states derived from a Yb$^+$ [Xe]$4f^{13}6s\sigma^2$ electronic configuration. These states have negative spin orbit parameters (e.g. $\zeta_{4f}<0$ for the $^2F^\circ_{7/2}$ and $^2F^\circ_{5/2}$ states of the Yb$^+$ ion \cite{kramidaNISTAtomicSpectra}), which will result in negative contributions to $\gamma$. Using the above estimates, and assuming that only the $\tilde{A}^2\Pi_{1/2}$ state and $4f^{13}6s\sigma^2$ states contribute to $\gamma$, gives $\gamma_{\nu_1=0}(f^{13}6s\sigma^2)=-0.03032$ cm$^{-1}$,$\gamma_{\nu_1=1}(4f^{13}6s\sigma^2)=-0.03106$ cm$^{-1}$, and $\gamma_{\nu_1=3}(4f^{13}6s\sigma^2)=-0.03261$ cm$^{-1}$. Here, $\gamma_{\nu_1}(4f^{13}6s\sigma^2)$ indicates the sum of the contributions to $\gamma_{\nu_1}$ from all electronic states derived from the $4f^{13}6s\sigma^2$ Yb$^+$ atomic electronic configuration.

The three spin-rotation parameters presented in Table \ref{tab: X300 A100 results} indicate that the vibrational dependence of $\gamma$ in the $\tilde{X}^2\Sigma^+$ state is linear with respect to the stretching vibration, $\nu_1$, with a slope of $\Delta\gamma/\Delta \nu_1\sim-0.001$ cm$^{-1}$. Additionally, the values of $\gamma$ increase in magnitude with increased stretching vibration. This is in contrast to the vibrational dependence of the estimated $\tilde{A}^2\Pi_{1/2}$ state contributions to $\gamma$, which decrease in magnitude with increasing stretching vibration at a linear rate which has a slope that is about five times smaller than the observed rate of change of the $\gamma$ values. This further indicates that the determined values of $\gamma$ in the $\tilde{X}^2\Sigma^+$ state cannot be accounted for by interactions solely with the $\tilde{A}^2\Pi_{1/2}$ state and other excited electronic states resulting from [Xe]$4f^{13}\sigma^2_{Yb^+(6s6p)}$ electronic configurations are involved. 

Recently, it was determined that low lying electronic states with a [Xe]$4f^{13}\sigma^2_{Yb^+(6s6p)}$ configuration are impacting the laser cooling and trapping of YbF \cite{zhangInnershellExcitationYbF2022}. It is expected that a similar situation will occur in YbOH. If leakage from the cooling cycle to these low-lying electronic states occurs at the 10$^{-5}$ level or more, it will need to be addressed in order to achieve laser slowing and magneto-optical trapping of YbOH.

In addition to determining the parameters for the $\xe$ state, the combined fit of the $\Ae-\xe$ obtained here and $\Ae-\xg$ data set from Ref.~\cite{steimleFieldfreeStarkZeeman2019} allowed a more accurate determination of the parameters of the $\Ae$ state. The values of the origin, $T_0$, rotational constant, $B$, and $\Lambda$-doubling centrifugal distortion parameter, $(p+2q)_D$, agree with the previously measured values~\cite{steimleFieldfreeStarkZeeman2019} and are more precisely determined, with the estimated errors a factor of 2 to 3 smaller. The determined value of the $\Lambda$-doubling parameter, $p+2q$, not only agrees with the previously determined value, but also is an order of magnitude more precise.

\section{\label{sec:Conclusion}Conclusion and Outlook}

We implement a method to measure higher-order repumping transitions for molecular laser cooling, and use it to perform new spectroscopic measurements in the excited streching modes of the YbOH ground and excited electronic states.  By increasing the molecular population in excited vibrational states via chemical production and using sensitive FM absorption, this method could be used to map out molecular repumping transitions without relying on complex optical cycling schemes.  Since the chemically-enhanced production method is generic to molecules with alkaline-earth (and similar) metals~\cite{jadbabaieEnhancedMolecularYield2020,vilasMagnetoopticalTrappingSubDoppler2022}, in particular many other laser-coolable species, this method should have wide utility. 

Simple improvements, such as using closed buffer gas cells to increase interaction time and length, and implementing multi-pass absorption paths, could increase the SNR to detect even weaker transitions out of other excited vibrational states. Additionally, the FM absorption technique we demonstrate here provides a path to perform direct spectroscopy on weak transitions involving low-lying electronic states, for example the [Xe]$4f^{13}\sigma^2_{Yb^+(6s6p)}$ states in Yb containing molecules that can cause optical cycling leakage. 

\vspace{3mm}

\begin{acknowledgments}
We would like to thank Greg Hall for all of his guidance, help, and advice when setting up the experimental FM setup and when modeling the FM lineshapes. We thank Timothy Steimle for his advice when developing the effective Hamiltonian model and fitting the spectrum. We would like to thank Phelan Yu and Ashay Patel for helpful discussions.

This work was supported by Heising-Simons Foundation grant numbers 2019-1193 and 2022-3361, and NSF CAREER award PHY-1847550.

\end{acknowledgments}

\bibliography{MyLibrary}

\appendix

\section{\label{Appx: FM model}FM lineshape model}

Experimentally, we measure the FM spectrum in the buffer gas cell where Doppler broadening dominates. Therefore, our absorption lineshape is best modeled as a Gaussian
\begin{equation}\label{eq: Gaussian absorption}
\delta(\omega) = A\, \exp\left( -\frac{(\omega-\omega_{res})^2}{2(\Gamma/2.355)^2} \right),
\end{equation}
where $A$ is the amplitude of the lineshape, $\omega_{res}$ is the linecenter, and $\Gamma$ is the full width at half maximum (FWHM). The dispersion lineshape can be calculated from the absorption lineshape using the Kramers-Kronig relationship \cite{forthommeArgonInducedPressureBroadening2013}
\begin{equation}
\phi(\omega)=\frac{1}{\pi}P \int_{-\infty}^{\infty} \frac{\delta(\omega')}{\omega'-\omega}d\omega',
\end{equation}
where $P$ is the Cauchy principal value. The Kramers-Kronig relation can also be represented as a Hilbert transform of the absorption line shape \cite{kingHilbertTransforms2009}. The Hilbert transform of a function $f(x)$ is given by
\begin{equation}
H(f)(y) = \frac{1}{\pi} P \int_{-\infty}^{\infty} \frac{f(x)}{y-x}dx.
\end{equation}
Therefore, given the absorption lineshape, $\delta(\omega')$ the dispersion line shape is given by the negative Hilbert transform of the absorption lineshape
\begin{equation}
\phi(\omega)=-H(\delta)(\omega).
\end{equation}
If we make the substitution $u=(\omega'-\omega_{res})/(\sqrt{2}\sigma)$ where $\sigma=\Gamma/2.355$ then $du=1/(\sqrt{2}\sigma)d\omega'$,  $\delta(\omega')=f(u)=A\, \exp(-u^2)$ and 
\begin{equation}
\begin{split}
\phi(\omega) &=\frac{1}{\pi} P \int_{-\infty}^{\infty} \frac{f(u)}{u-\frac{\omega-\omega_{res}}{\sqrt{2}\sigma}}du \\
\\
&= \frac{1}{\pi} P \int_{-\infty}^{\infty} \frac{-A\times \exp(-u^2)}{t-u}du \\
\\
&= -A\, H[\exp(-u^2)](t),
\end{split}
\end{equation}
where $t=(\omega-\omega_{res}/(\sqrt{2}\sigma)$. The Hilbert transform of $f(u)=\exp(-u^2)$ is known and is related to the Dawson function \cite{kingHilbertTransforms2009a,kingHilbertTransforms2009}
\begin{equation}
H[\exp(-u^2)](t)=\frac{2}{\sqrt{\pi}}F(t),
\end{equation}
where $F(t)$ is the Dawson function. Therefore, the dispersion lineshape is given by
\begin{equation}\label{eq: dispersion lineshape}
\phi(\omega)=-A\frac{2}{\sqrt{\pi}}F\left(\frac{(\omega-\omega_{res})}{\sqrt{2}(\Gamma/2.355)}\right).
\end{equation}
The dispersion lineshape given in Eq. \ref{eq: dispersion lineshape} is convenient for numerical modeling since the Dawson function is a built-in function in several programming languages.

Ultimately, we want to model a true absorption spectrum which will contain an arbitrary superposition of Gaussian lineshapes
\begin{equation}\label{eq: absorption total}
\begin{split}
\delta_{tot}(\omega) &= \sum_{i}\delta_i(\omega) \\
&= \sum_{i} A_i\, \exp\left(-\frac{(\omega-\omega_i)^2}{2(\Gamma_i/2.355)}\right),
\end{split}
\end{equation}
where $i$ denotes the absorption lineshape due to the $i^\textrm{th}$ transition, and $\omega_i$ is the resonance frequency of the $i^\textrm{th}$ transition. Therefore, the total dispersion lineshape due to the combination of multiple transitions is given by
\begin{equation}\label{eq: dispersion total}
\begin{split}
\phi_{tot}(\omega) &=\frac{1}{\pi}P\int_{-\infty}^{\infty} \frac{\delta_{tot}(\omega')}{\omega'-\omega}d\omega \\
\\
& = \sum_i A_i\frac{2}{\sqrt{\pi}}F\left(\frac{(\omega-\omega_i)}{\sqrt{2}(\Gamma_i/2.355)}\right).
\end{split}
\end{equation}

With Eq. \ref{eq: absorption total} and \ref{eq: dispersion total}, any arbitrary absorption and dispersion line shape can be modeled.

In the weak absorption limit, the FM lineshapes to second order due to absorption and dispersion are given by \cite{forthommeArgonInducedPressureBroadening2013}
\begin{equation}\label{eq: A FM 2nd}
\begin{split}
A_{FM}(\omega) =& J_0(M)J_1(M)\left[\delta(\omega-\omega_m)-\delta(\omega+\omega_m)\right] \\
\\
&+ J_1(M)J_2(M)\\
&\times\left[ \delta(\omega-2\omega_m)-\delta(\omega+2\omega_m)\right. \\
&\left. +\delta(\omega-\omega_m)-\delta(\omega+\omega_m) \right],
\end{split}
\end{equation}
and
\begin{equation}\label{eq: D FM 2nd}
\begin{split}
D_{FM}(\omega) =& J_0(M)J_1(M)\\
&\times\left[ \phi(\omega-\omega_m) + \phi(\omega+\omega_m) -2\phi(\omega) \right]\\
\\
& + J_1(M)J_2(M)\\
&\times\left[ \phi(\omega-2\omega_m) + \phi(\omega+2\omega_m)\right.\\
&\left. -\phi(\omega-\omega_m)-\phi(\omega+\omega_m) \right],
\end{split}
\end{equation}respectively. Here, $\delta(\omega)$ is the total absorption lineshape, $\phi(\omega)$ is the total dispersion lineshape, $J_n(M)$ is the Bessel function of order $n$, $M$ is the modulation depth ($M=$0.84), $\omega$ is the carrier frequency of the laser, and $\omega_m$ is the modulation frequency applied to the laser by the EOM. The resulting dc signals following the I and Q demodulator will depend on the phase angle, $\theta$, the phase difference between the two paths from the rf oscillator to the I and Q demodulator. At an arbitrary value of $\theta$, the output in-phase, $I_{FM}$, and in-quadrature, $Q_{FM}$, FM signals will be sine and cosine weighted mixtures of the absorption and dispersion FM signals \cite{hallTRANSIENTLASERFREQUENCY2000}
\begin{equation}\label{eq: I FM}
I_{FM}(\omega)= \cos\theta A_{FM}(\omega) + \sin\theta D_{FM}(\omega),
\end{equation}and
\begin{equation}\label{eq: Q FM}
Q_{FM}(w) = \sin\theta\, A_{FM}(\omega)-\cos\theta \,D_{FM}(\omega).
\end{equation}

Given a set of transitions (each with transition frequency $\omega_i$, width $\Gamma_i$, and amplitude $A_i$) and a phase angle, $\theta$, the  $I_{FM}(\omega)$ and $Q_{FM}(\omega)$ lineshapes are calculated using Eq. \ref{eq: I FM} and \ref{eq: Q FM} respectively, where $A_{FM}(\omega)$ is given by Eq. \ref{eq: A FM 2nd}, $D_{FM}(\omega)$ is given by Eq. \ref{eq: D FM 2nd}, $\delta(\omega)$ is given by Eq. \ref{eq: absorption total}, and $\phi(\omega)$ is given by Eq. \ref{eq: dispersion total}. 

Ultimately, we want to perform a simultaneous fit of our modeled $I_{FM}$ and $Q_{FM}$ FM lineshapes to the measured $I_{FM}$ and $Q_{FM}$ data. This is accomplished with a non-linear least squares optimization~\cite{NonLinearLeastSquaresMinimization} which takes the FM data and initial guesses for the phase angle and the parameters of each transition present (the transition frequency, width, and amplitude of each transition) as inputs. The optimization works to minimize the set of residuals provided to it. To accomplish the simultaneous fit the following residual function was used
\begin{equation}
R(\omega) = \sqrt{\left[ I_{calc}(\omega)-I_{data}(\omega)\right]^2+\left[ Q_{calc}(\omega)-Q_{data}(\omega)\right]^2}
\end{equation}
where $I_{calc}(\omega)$ ($Q_{calc}(\omega)$) is the calculated value of $I_{FM}$ ($Q_{FM}$) at the frequency $\omega$ and $I_{data}(\omega)$ ($Q_{data}(\omega)$) is the measured value of $I_{FM}$ ($Q_{FM}$) at the frequency $\omega$. The sum of the squares of the individual I and Q residuals as opposed to just the sum of the I and Q residuals was used to prevent the residual from taking on inaccurately small values due to a cancellation resulting from the I and Q residuals being opposite in sign. In the fit the phase angle $\theta$ and the lineshape parameters $\omega_i$, $\Gamma_i$, and $A_i$ are floated. Any arbitrary number of transitions can be fit by the algorithm. For isolated lines we find that the linecenters extracted from the fit exactly match our zero crossing measurements and have equivalent or smaller errors.  

\section{Lines and Fit Residuals}

\begin{table*}[ht]

\caption{\label{tab: X300 - A100 FM lines} The transition wavenumbers and assignments for the $\Ae-\xe$ band of $^{174}$YbOH measured with in-cell FM spectroscopy. Also presented are the differences between the observed (Obs.) and calculated (Calc.) transition wavenumbers. The calculated values were obtained using the optimized parameters from the combined fit of both the $\Ae-\xe$ and $\Ae-\xg$ \cite{steimleFieldfreeStarkZeeman2019} data sets. Here $p$ indicates the parity of the molecular state.}

\begin{ruledtabular}
\begin{tabular}{c c c c c c c c c c}

Lines & $N''$, $J''$, $p$ & $J'$, $p$ & Obs.  &  Obs.$-$ Calc. &Lines & $N''$, $J''$, $p$ & $J'$, $p$ & Obs.  &  Obs.$-$ Calc.  \\
& & & (cm$^{-1}$)& (MHz)& & & & (cm$^{-1}$)& (MHz)\\
\hline\\

$^OP_{12}$&2, 1.5, + &0.5, - & 16335.7340&15 &$^QQ_{11}$& 0, 0.5, +&0.5, -&16337.1876&21\\
$^PQ_{12}$& 2, 1.5, +&1.5, -&16337.2948&-4 & &1, 1.5, -&1.5, +&16337.2004&4\\
& 3, 2.5, - & 2.5, + & 16337.3806 & 3 & &2, 2.5, +&2.5, -&16337.2384&5\\
& 4, 3.5, + & 3.5, - & 16337.4909 & 1 & &3, 3.5, -&3.5, +&16337.3004&-11\\
& 5, 4.4, -& 4.5,+& 16337.6261& -6 & &4, 4.5, +&4.5, -&16337.3875&-8\\
& 6, 5.5, + & 5.5, -& 16337.7863& -11 & &5, 5.5, -&5.5, +&16337.4993&2\\
& 7, 6.5,-& 6.5,+ &16337.9717& -10 & &6, 6.5, +&6.5, -&16337.6347&-10\\
& 8, 7.5, +& 7.5, - & 16338.1833&23 & &7, 7.5, -&7.5, +&16337.7948&-11\\
$^PP_{11}$& 1, 1.5, -&0.5, +&16337.2431&9 & &8, 8.5, +&8.5, -&16337.9793& -9\\
& 2, 2.5, +& 1.5, -& 16337.3095& 6 & &9, 9.5, -&9.5, +&16338.1895&38\\
&3, 3.5, -&2.5, +&16337.4005&-4 & $^QR_{12}$& 1, 0.5, -& 0.5, +&16337.1922& 17\\
& 4, 4.5, +&3.5, -& 16337.5166&-7 & &2, 1.5, +&2.5, -&16337.2240&4\\
& 5, 5.5, -& 4.5, +&16337.6577&-7 & &3, 2.5, -&3.5,+& 16337.2801& -16\\
&6, 6.5, +&5.5, -&16337.8237&-11 & &4, 3.5, +&4.5, -&16337.3619& 1\\
&7, 7.5, -&6.5 +&16338.0150&-6 & &5, 4.5, -&5.5, +&16337.4674&-5\\
&8, 8.5, +&7.5, -&16338.2314&-1 & &6, 5.5, +&6.5, -&16337.5974&-7\\
$^RR_{11}$&0, 0.5, +&1.5, -&16338.7486&9 & &8, 7.5, +&8.5, -&16337.9305&-5\\
&1, 1.5, -& 2.5, +& 16339.8026&-6 & &Unassigned& &16338.6793&\\
&            &          &                     &    & &Unassigned& &16338.7399&\\
&            &          &                     &    & &Unassigned& &16339.5994&\\
\end{tabular}
\end{ruledtabular}
\begin{flushleft}
RMS of combined fit with $\Ae-\xg$ data \cite{steimleFieldfreeStarkZeeman2019}: 25 MHz (0.00084 cm$^{-1}$)\\
\end{flushleft}
\end{table*}

\begin{table*}[ht]

\caption{\label{tab: X000 - A100 lines} The transition wavenumbers and assignments for the $\Ae-\xg$ band of $^{174}$YbOH from Ref \cite{steimleFieldfreeStarkZeeman2019} Also presented are the differences between the observed (Obs.) and calculated (Calc.) transition wavenumbers. The calculated values were obtained using the optimized parameters from the combined fit of both the $\Ae-\xe$ and $\Ae-\xg$ data. Here $p$ indicates the parity of the molecular state.}

\begin{ruledtabular}
\begin{tabular}{c c c c c c c c c c}

Lines & $N''$, $J''$, $p$ & $J'$, $p$ & Obs.  &  Obs.$-$ Calc. &Lines & $N''$, $J''$, $p$ & $J'$, $p$ & Obs.  &  Obs.$-$ Calc.  \\
& & & (cm$^{-1}$)& (MHz)& & & & (cm$^{-1}$)& (MHz)\\
\hline\\

$^OP_{12}$&2, 1.5, +&  0.5, -&  17906.3830 & 35 & $^RR_{11}$&2, 2.5, +& 3.5, -& 17911.5217& -34\\
&3, 2.5, -&  1.5, +& 17905.4024 & 15 & &3, 3.5, -& 4.5, +& 17912.5999& 10\\
&4, 3.5, +&  2.5, -& 17904.4382 & 8 & &4, 4.5, +& 5.5, -& 17913.6912& -42\\
&5, 4.5, -& 3.5, +& 17903.4897 & -7 & &5, 5.5, -& 6.5, +& 17914.8019& -2\\
&6, 5.5, +& 4.5, -& 17902.5568 & -29 & $^QQ_{11}$&0, 0.5, +& 0.5, -& 17907.8571& 15\\
$^PQ_{12}$&1, 0.5, -& 0.5, +& 17907.8990& 22 & &1, 1.5, -& 1.5, +& 17907.8603& 15\\
&2, 1.5, +& 1.5, -& 17907.9442& 29 & &2, 2.5, +& 2.5, -& 17907.8793& 10\\
&3, 2.5, -& 3.5,+ & 17908.0046& 6 & &3, 3.5, -& 3.5, +& 17907.9132& -25\\
&4, 3.5, +& 3.5, -& 17908.0810& -24 & &4, 4.5, +& 4.5, -& 17907.9648& -4\\ 
&5, 4.5, -& 4.5, +& 17908.1752& -8 & &5, 5.5, -& 5.5, +& 17908.0321& 15\\
&6, 5.5, +& 5.5, -& 17908.2853& -5 & &6, 6.5, +& 6.5, -& 17908.1130& -30\\
&7, 6.5, -& 6.5, +& 17908.4137& 56 & &7, 7.5, -& 7.5, +& 17908.2113& -23\\
&8, 7.5, +& 7.5, -& 17908.5533& -40 & &8, 8.5, +& 8.5, -& 17908.3256& -3\\
&9, 8.5, -& 8.5, +& 17908.7132& -21 & &9, 9.5, -& 9.5, +& 17908.4542& -21\\
&10, 9.5, +& 9.5, -& 17908.8897& 0.3 & &10, 10.5, +& 10.5, -& 17908.5992& -10\\
&11, 10.5, -& 10.5, +& 17909.0810&-31 & &11, 11.5, -& 11.5, +& 17908.7604& 24\\
&12, 11.5, +& 11.5, -& 17909.2912& 6& &12, 12.5, +& 12.5, -& 17908.9357& 22\\
$^PP_{11}$&1, 1.5, -& 0.5, +& 17907.9028& 15 & &13, 13.5, -& 13.5, +& 17909.1231& -75\\
&2, 2.5, +& 1.5, -& 17907.9508&24 & &14, 14.5, +& 14.5, -& 17909.3319&  14\\
&3, 3.5, -& 2.5, +& 17908.0145& 19 & &15, 15.5, -& 15.5, +& 17909.5545& 63\\
&4, 4.5, +& 3.5, -& 17908.0937& -8 & $^QR_{12}$&1, 0.5, -& 1.5, +& 17907.8571& 41\\
&5, 5.5, -& 4.5, +& 17908.1901& -7 & &2, 1.5, +& 2.5, -& 17907.8725& 9\\
&6, 6.5, +& 5.5, -& 17908.3042& 35 & &3, 2.5, -& 3.5, +& 17907.9036& -29\\
&7, 7.5, -& 6.5, +& 17908.4335& 43 & &4, 3.5, +& 4.5, -& 17907.9535& 22\\
&8, 8.5, +& 7.5, -& 17908.5762& -40 & &5, 4.5, -& 5.5, +& 17908.0175& 23\\
&9, 9.5, -& 8.5, +& 17908.7388& -21 & &6, 5.5, +& 6.5, -& 17908.0953& -35\\
&10, 10.5, +& 9.5, -& 17908.9165& -43 & &7, 6.5, -& 7.5, +& 17908.1908& -31\\
&11, 11.5, -& 10.5, +& 17909.1120& -28 & &8, 7.5, +& 8.5, -& 17908.3035& 21\\
&12, 12.5, +& 11.5, -& 17909.3242& -9 & &9, 8.5, -& 9.5, +& 17908.4278& -45\\
&13, 13.5, -& 12.5, +& 17909.5545& 53 & &10, 9.5, +& 10.5, -& 17908.5719& 18\\
&14, 14.5, +& 13.5, -& 17909.7992& 47 & &11, 10.5, -& 11.5, +& 17908.7297& 30\\
&                 &            &                   &      & &12, 11.5, +& 12.5, -& 17908.9024& 29\\
&                 &            &                   &      & &13, 12.5, -& 13.5, +& 17909.0869& -77\\
&                 &            &                   &      & &14, 13.5, +& 14.5, -& 17909.2931& 13\\
\end{tabular}
\end{ruledtabular}
\begin{flushleft}
RMS of combined fit with $\Ae-\xe$ data: 25 MHz (0.00084 cm$^{-1}$)\\
\end{flushleft}
\end{table*}

\end{document}